# How to Read The Snowmass White Papers
## on
## *Power Dynamics in Physics*
## *Informal Socialization in Physics Training*
## and
## *Policing and Gatekeeping in STEM*

Apriel K Hodari,[1] Shayna B Krammes[1]
Chanda Prescod-Weinstein,[2] Brian D Nord,[3] Jessica N Esquivel,[3] Kétévi A Assamagan[4]

[1]Eureka Scientific Inc, Oakland, CA 94602
[2]University of New Hampshire, Durham, NH 03824
[3]Fermi National Accelerator Laboratory, Batavia, IL 60510
[4]Brookhaven National Laboratory, Upton, NY 11973

The **Community Engagement Frontier** presents this set of three white papers, as part of Snowmass 2021. These papers address critical issues—*Power Dynamics in Physics, Informal Socialization in Physics Training*, and *Policing and Gatekeeping in STEM*—that make significant impacts on the experiences of the people who work in and learn particle physics (Hodari et al., 2022a, 2022b, 2022c).

In this introductory document, we present crosscutting concepts that appear in each paper, and some advice on how to manage readers' responses to the contents. We expect that you will learn something new here. We hope that whatever you encounter, you will be energized to increase justice in this discipline we all love.

<u>CROSSCUTTING CONCEPTS</u>
*Defining the Lie*
In *Begin Again*, scholar and author Eddie S Glaude, Jr defined the, "set of practices that, taken together, constitute … *the lie*". He explained (2020, p. 8):

> *The lie is more properly several sets of lies with a single purpose. If what I have called the 'value gap' is the idea that in America white lives have always mattered more than the lives of others, then the lie is a broad and powerful architecture of false assumptions by which the value gap is maintained. These are narrative assumptions that support the everyday order of American life, ...*

> *One set of lies debases black people … [who are] essentially inferior, less human than white people, and therefore deserving of their particular station in American life.*

> *Another constituent part of the lie involves lies about American history and about the trauma that America has visited throughout that history on people of color … America is fundamentally good and innocent, its bad deeds dismissed as mistakes corrected on the way to 'a more perfect union.'… The genocide of native peoples, slavery, racial*

> *apartheid, Japanese internment camps, and the subordination of women reveal that our basic creed that 'all men are created equal' was a lie, at least in practice. ... Each moment represented a profound revelation about who we were as a country—just as the moments of resistance against them said something about who we aspired to be.*
>
> *But the lie's most pernicious effect when it comes to our history is to malform events to fit the story whenever America's innocence is threatened by reality.*

Across this set of papers, when we refer to *the lie*, we include the full set of oppressions—racism, sexism, classism, homophobia, transphobia, ableism, ageism, fatphobia—that are supported by practice, sometimes policy and law; are reified by American culture; and thus seep into our disciplinary culture.

### *Physics as "A Culture of No Culture"*

One of the most pervasive elements of our disciplinary culture is the belief that physics, the purity of the science itself, creates a space that remains uninfluenced by any outside factors. Sharon Traweek described and defined the phrase "a culture of no culture" in 1988 (Traweek):

> *I have never met a high energy physicist who would entertain for a moment the question of whether electrons 'exist' or not; and I can sympathize with that, for unlike some of my more reflexivist colleagues, I find it appropriate to assume that physicists exist. Unlike most physicists, though, I do recognize the importance of the question, in a less abrupt form: where do the social categories of physicist and physics community of physics culture exist? I mean this book to address that question. I have presented an account of how high energy physicists construct their world and represent it to themselves as free of their own agency, a description, as thick as I could make it, of an extreme culture of objectivity: a culture of no culture, which longs passionately for a world without loose ends, without temperament, gender, nationalism, or other sources of disorder—for a world outside human space and time.*

This is not a special form of mental gymnastics only performed by physicists. "A culture of no culture" was later extended to characterize sexist "(de)mentoring" of women across multiple STEM disciplines and objectivity beliefs in medicine (Subramaniam & Wyer, 1998; Taylor, 2003). Key features of cultures described by this term are deeply engrained belief in objectivity as the supreme value, and the pervasive oppression that belies that belief. Objectivity is a thinly-veiled mask for the cruelty that maintains pecking order, what Traweek described in this passage about hierarchy and cultural commonality that eclipses all others:

> *The members of the particle physics community are firmly committed to the international, supracultural image of science* [(Restivo & Vanderpool, 1974; Storer, 1974)]. *Particle physicists from anywhere in the world are fond of remarking that they have more in common with each other than with their next-door neighbors. All of these physicists consider themselves of an intellectual elite, perhaps **the** intellectual elite, because they believe particle physics works alone at the frontiers of human knowledge.*

The elitism, the hierarchy that reifies it, and the oppression that enforces it mirror the consequences of *the lie*. It is therefore not surprising that those who thrive within existing physics culture—overwhelmingly white men—see the culture of no culture as normal, and thereby reproduce oppression, despite individual intention, earnest sincerity, or personal belief.

### W̲HAT YOU T̲HINK AND B̲ELIEVE

Author and somatic therapist Resmaa Menakem starts his book *My Grandmother's Hands* with direct statements about how readers may react to the content based on what they believe about systemic oppression (2017). Based on the reader's beliefs, Menakem offers insights into how the book may challenge or confirm their beliefs. For some readers, Menakem suggests that engaging the material may trigger their trauma reflexes and cause them additional pain, so he recommends they not read the book.

Many readers, depending on how many privileged identities they embody, will find these white papers challenging to read and process. For those with multiple undervalued identities, they may find confirmation of their experiences. Whatever your experience, we have worked hard to offer knowledges and suggestions grounded in solid research and the lived experiences of people who hold different perspectives. The primary motivation is to make these offerings in an accessible form, rich with resources that you can use to deepen your understandings.

### W̲ATCH Y̲OUR B̲ODY

Another set of suggestions Menakem makes regards how readers' bodies may experience the ideas he presents (2017). We summarize his suggestions regarding the three papers presented here. Fundamentally, our intent is to offer insights and resources. However, we acknowledge, based on your positionality, that you may experience bodily discomfort—reflexive constriction, defensive thoughts, a "sudden shock of recognition or understanding"(DiAngelo & Menakem, 2020; Menakem, 2017) . You may also experience a rush of emotional energy—joy, anger, outrage, clarity, rightness—or some combination of all of these responses.

As Menakem recommends, we suggest you allow these responses to moving through you without holding on to them. Once you have done that, we suggest you accept the opportunity Glaude offers in *Begin Again* (2020). Ask yourself, with the goal of increasing justice for all members of our disciplinary community, "What do these [*perhaps new* knowledges] commend to us to do?"

### W̲HAT'S IN A N̲AME

We acknowledge that it is becoming increasingly common to capitalize "Black" but not "white" when writing about race. To capitalize "Black" is to interrupt the historic positioning of white over black. Additionally, the choice to *not* capitalize white is significant. "White" is a racial category that holds substantial significance, but it is often written in all lowercase to avoid aligning with white supremacist groups (Bauder, 2020; DiAngelo, 2021, p. xix). For the sake of simplicity, in this set of papers we will not be capitalizing black, white, or brown. We will also use the term African American interchangeably with black, depending on context and who we are quoting or citing.

Whenever possible, we will use specific countries and tribal names when talking about a person or group. In cases when the specifics are not known to us, we will use more broad terms such as Indigenous, Asian American, etc. We will also use the term "BIPOC" in this set of papers, which refers broadly to all people of color. "BIPOC" stands for black, Indigenous, and people of color. It recognizes the particular significance of black and Indigenous people in the US, given that these two groups, broadly speaking, have faced the most severe and enduring consequences of racism and white supremacy. In our writing, although not grammatically correct, we usually use the term "BIPOC people" because it is more humanizing (DiAngelo, 2021, p. xviiii).

Latin America is an expansive geographical and cultural region of the world, and we acknowledge that many people from this region identify more with their country of origin than with the Latin American collective. This is also often true for immigrants to the US, regardless of generation. However, people from Latin America are typically grouped together under one label once they are in the United States. Government documents use the term "Hispanic" to describe this group, but this term centers the history of colonization and is losing popularity. Another common label is "Latino/a." When speaking collectively about this diverse group of people, we will use the term "Latinx" because it is both gender-neutral and disrupts the gender binary (Balta, 2021; Morales, 2018). The Spanish language is heavily gendered and many people, from scholars to politicians to the general public, continue to wrestle with how best to navigate this (Paz, 2021; Peñaloza, 2020). However, Latinx is the most well-known option at the time of writing.

On occasion, our language will differ from what we have outlined here, either when quoting verbatim or directly citing a source that uses a specific term. For example, "Hispanic" is still commonly used in studies and will appear in our writing at times. The only exception to this will be when quotations include identifiable information or racial slurs, in which case we will make the appropriate and necessary edits to remove them.

### READING FOR WHAT'S MISSING

We invite readers to pay attention to what the data doesn't tell us. For example, often charts and documents about underrepresented minorities in STEM simply leave Indigenous Americans off statistical accounts entirely. The argument for doing this is that Indigenous Americans do not participate in STEM at high enough rates for statistical analyses to be possible. Of course, this fact alone is data, and that information should be made visible. Because not all researchers have committed to the visibility of this severe underrepresentation and inclusion, it is on us, the readers, to notice it.

Another example is datasets about women that do not disaggregate by race, thus making invisible the intersectional forces that shape the lives of women of color. When they do disaggregate, there is a tendency to treat black and Hispanic or Latinx, for example, as mutually exclusive identities, erasing the experiences of Afro-Latinx people. We run into the same problem when analyses about gender assume a gender binary, excluding nonbinary and gender expansive people.

In some sense, this encouragement we are offering is to be curious. Whose stories are told and which ones are missing? And why might those stories be hidden from view? How might paying attention to what's not in the data or the data analysis strengthen our capacity to confront challenges in our communities, dynamically, as they unfold?

### ACKNOWLEDGEMENTS
The authors acknowledge the Heising-Simons Foundation Grant #2020-2374 which supported this work. We also thank all the non-author organizers for the 2021 A Rainbow of Dark Sectors conference: Regina Caputo, Djuna Croon, Nausheen Shah, and Tien-Tien Yu. We additionally thank Risa Wechsler for her organizational support.